\documentclass{nature2}
\usepackage{amsmath}
\usepackage{graphicx}
\usepackage{tabularx}
\usepackage{longtable}
\usepackage{caption}

\title{The use of double-mode RR Lyrae stars as robust distance and metallicity indicators}

\author{Xiaodian Chen$^{1,2,3,4,5}$, Jianxing Zhang$^{1,3,5}$, Shu Wang$^{1,3,4}$ and Licai Deng$^{1,3,4}$}

\begin{document}

\maketitle

\begin{affiliations}
 \item CAS Key Laboratory of Optical Astronomy, National Astronomical Observatories, 
   Chinese Academy of Sciences, Beijing, 100101, China
 \item Institute for Frontiers in Astronomy and Astrophysics, Beijing Normal University,  Beijing 102206, China
 \item School of Astronomy and Space Science, University of the Chinese Academy of Sciences, Beijing, 100049, China
 \item Department of Astronomy, China West Normal University, Nanchong, 637009, China
 \item These authors contributed equally: Xiaodian Chen, Jianxing Zhang.
\end{affiliations}

\begin{abstract}
RR Lyrae stars are one of the primary distance indicators for old stellar populations such as globular clusters, dwarf galaxies and galaxies. Typically, fundamental-mode RR Lyr stars are used for distance measurements, and their accuracy is strongly limited by the dependence of absolute magnitudes on metallicity, in both the optical and infrared bands. Here, we report the discovery of a period–(period ratio)–metallicity relation for double-mode RR Lyr stars, which can predict metallicity as accurately as the low-resolution spectra. With theoretical and observational evidence, we propose that the period– luminosity relation of double-mode RR Lyr stars is not affected by
the metallicity. Combining the Large Magellanic Cloud distance and Gaia parallaxes, we calibrate the zero point of the period–luminosity relation to an error of 0.022 mag, which means that in the best case double-mode RR Lyr stars can anchor galaxy distances to an accuracy of 1.0\%. For four globular clusters and two dwarf galaxies, we obtain distances using double-mode RR Lyr stars with a distance accuracy of 2–3\% and 1–2\%, respectively. With future telescopes such as the China Space Station Telescope and the Vera C. Rubin Observatory, double-mode RR Lyr stars will be established as an independent distance ladder in the near-field universe.
\end{abstract}

\begin{figure}
\begin{center}
\includegraphics[width=160mm]{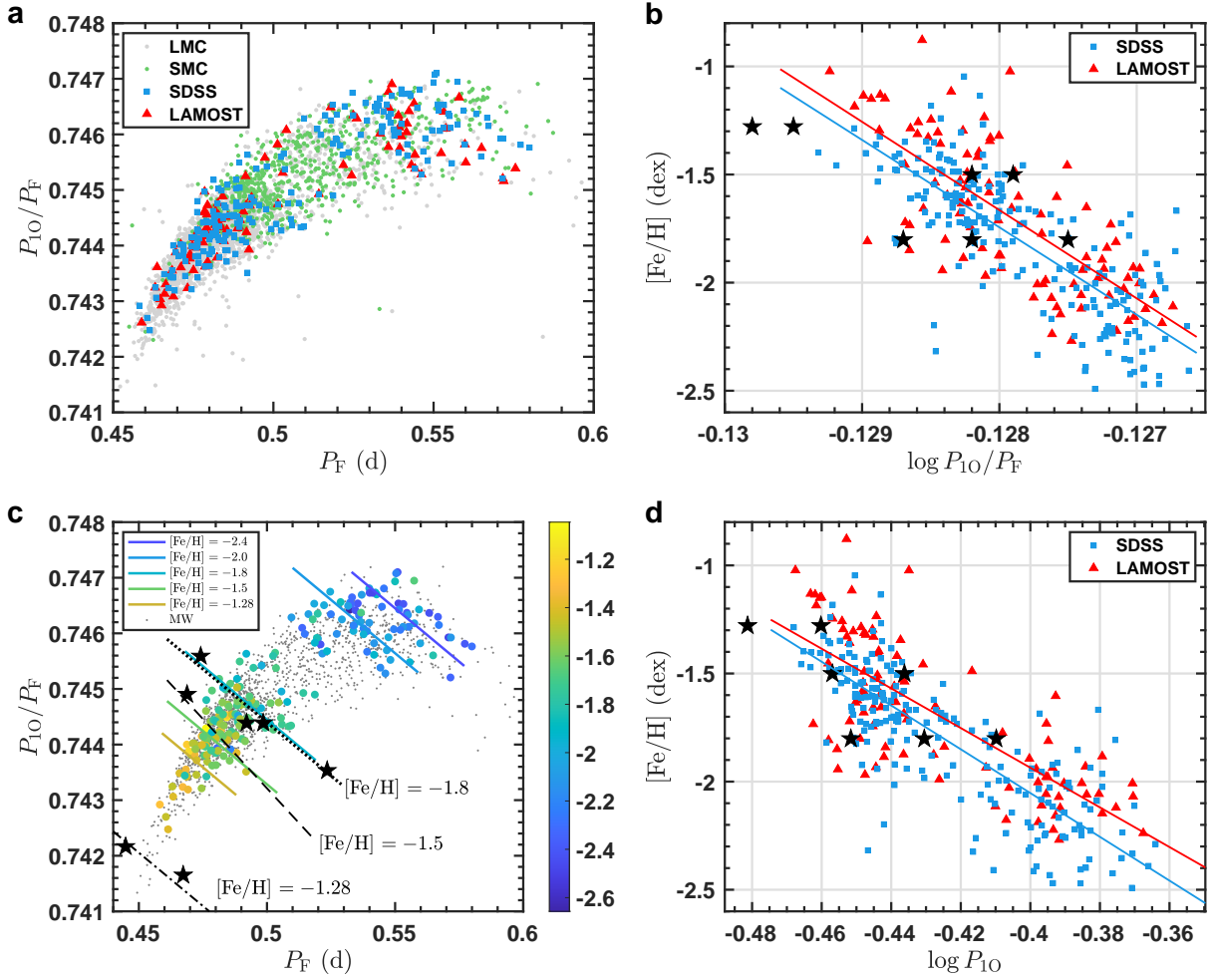}
\caption{\label{Fig1}{\bf Period ratio--period diagram (Petersen diagram), period ratio--metallicity diagram, and period--metallicity diagram.} {\bf a}, The light green and light gray dots indicate the loci of the SMC and LMC RRd stars on the Petersen diagram. 207 and 96 RRd stars with SDSS and LAMOST metallicities are shown as blue squares and red triangles, respectively. $P_{\rm F}$ and $P_{\rm 1O}$ denote the fundamental period and first-overtone period. {\bf b}, The blue squares and red triangles show the period ratio--metallicity relations for RRd stars based on SDSS and LAMOST metallicities, respectively. The best-fit lines are in the same color. The black stars in panels b, c, d show the grid points of theoretical models from ref.\cite{Marconi15}. {\bf c},  On the Petersen diagram, the RRd stars are represented by solid circles of different colors according to their SDSS metallicities. Milky Way's RRd stars are indicated by black dots as the background. The fitted equal-metallicity lines are represented by different solid lines, respectively, and the colors are kept consistent with the color bars. The theoretical models (ref.\cite{Marconi15}) of the three different metallicities are represented by dotted, dashed and dot-dashed lines. {\bf d}, Period--metallicity diagram. Symbols are as panel b.} 
\end{center}
\end{figure}

RR Lyrae stars simultaneously pulsating at two different periods are classified as double-mode RR Lyr stars (RRd). The majority of them are classical RRd stars, characterized by the presence of the radial fundamental and first-overtone modes, where the first-overtone mode is usually the dominant mode. The default RRd stars in this paper are the classical RRd stars. We collected 1021 Galactic RRd stars from Gaia DR3\cite{Clementini2022}, 2083 and 674 RRd stars belonging to the Large Magellanic Cloud (LMC) and the Small Magellanic Cloud (SMC) from the Optical Gravitational Lensing Experiment (OGLE) database\cite{Soszynski2016}, and plotted the period--period ratio diagram (Petersen diagram\cite{Petersen1973}) in Fig. \ref{Fig1}a and Fig. \ref{Fig1}c. At each period, the RRd stars in different galaxies have a consistent period ratio distribution. Theory\cite{Popielski2000, Marconi15} and observations\cite{Bragaglia2001, Coppola15} suggest that metal-rich RRd stars have a shorter fundamental period ($P_{\rm F}$) and a smaller period ratio ($P_{\rm 1O}/P_{\rm F}$) compared to metal-poor RRd stars. We found that SMC RRd stars have longer mean periods and larger mean period ratios than the LMC and Galactic RRd stars (see Extended Data Fig. 1). Alternatively, the distribution of periods and period ratios of LMC and Galactic RRd stars is wider.

We cross-matched Gaia DR3's RRd stars with all available spectroscopic observations and found 68 and 32 with metallicities from the Sloan Digital Sky Survey\cite{Eisenstein2011} (SDSS) and the Large Sky Area Multi-Object Fiber Spectroscopic Telescope\cite{Cui2012} (LAMOST), respectively. Based on the Zwicky Transient Facility (ZTF) DR14 photometry, we additionally discovered twice the number of RRd stars that have spectral parameters. Our final sample contains 207 and 96 RRd stars with metallicity measurements from SDSS and LAMOST, respectively. These metallicities were measured from low-resolution spectra ($R\sim2000$) with an external error of 0.13-0.19 dex. For either the SDSS or LAMOST metallicities, we found an intuitive linear relationship between the metallicity and the period ratio (Fig. \ref{Fig1}b) or period (Fig. \ref{Fig1}d). The best-fit period--period ratio--metallicity relations are shown in Equ. \ref{equ1}. $\sigma$ is the dispersion of the relations, while $R^2$ is the coefficient of determination. The use of logarithmic period ratios and periods facilitates comparison with theoretical results. We also adopted a mean period ratio of 0.745 and a mean period of 0.37 days as the zero point so that the intercept can directly present the metallicity. The relations for SDSS and LAMOST RRd stars are consistent with each other if uncertainties are taken into account. The main contribution of dispersion is the metallicity uncertainty of the low-resolution spectra. When we selected a better sample based on the internal error of metallicity, the dispersion of the period-period ratio-metallicity relation gradually decreases as the criterion becomes tighter. The best case is when limiting $\sigma_{\rm [Fe/H]}<0.04$ dex, based on 56 SDSS RRds, we obtained a dispersion of 0.13 dex. By error comparison (also see method), we found that the metallicity estimated using the period and the period ratio of RRd stars can be as accurate as the low-resolution spectra. The subsequent discussions in this paper are based on the SDSS RRd stars considering the larger number and the smaller metallicity uncertainty. 

\begin{equation}\label{equ1}
 \begin{aligned}
   {\rm [Fe/H]_{SDSS}} = & -(173\pm38)(\log \frac{P_{\rm 1O}}{P_{\rm F}}-\log0.745)-(6.62\pm0.88)(\log P_{\rm 1O}-\log0.37) \\
   &-(1.76\pm0.01),  \sigma=0.16 \, {\rm dex}, R^2=0.74,\\
   {\rm [Fe/H]_{LAMOST}} = & -(184\pm70)(\log \frac{P_{\rm 1O}}{P_{\rm F}}-\log0.745)-(5.69\pm1.50)(\log P_{\rm 1O}-\log0.37) \\
   &-(1.68\pm0.03),  \sigma=0.21 \, {\rm dex}, R^2=0.60,\\
 \end{aligned}
\end{equation}

 Due to an approximate linear correlation ($R^2=0.70$) between period and period ratio, the period--period ratio--metallicity relation can be simplified to period--metallicity relation or period ratio--metallicity relation. The dispersion of the linear relation between $\log P_{\rm F}$ and $\log P_{\rm 1O}/P_{\rm F}$ is $\sigma_{\log P_{\rm 1O}/P_{\rm F}}=0.0003$ (corresponding to a metallicity dispersion of 0.05 dex from Equ. \ref{equ1}). In terms of accuracy, the period--metallicity relation is closer to the period--period ratio--metallicity relation. Equ. \ref{equ11} shows the determined period--metallicity relations for SDSS RRd stars, using the same sample as Equ. \ref{equ1}. These relations are important for the optimization of the theoretical model and the use of RRd stars for high-precision distance measurements. In Fig. \ref{Fig1}, we compared our relations with the theoretical model of RRd stars\cite{Marconi15}. Here we assumed that the metal abundance of the Sun is $Z = 0.019$ and that all heavy elements vary by the same factor in different RRd stars. The black star symbols are 7 theoretical grid points  (Zero-Age-Horizontal-Branch luminosity level) located near the observed sequences on the Petersen diagram. The general trends of theoretical grid points and observed period ratio--metallicity relations (Fig. \ref{Fig1}b) and period--metallicity relations (Fig. \ref{Fig1}d) are consistent. However, the uncertainty of the period ratio calculated by the nonlinear pulsation theory (see Fig. \ref{Fig1}b) is 1-2 orders of magnitude larger than that of the observations ($\sigma_{P_{\rm 1O}/P_{\rm F}}=2\times10^{-5}$). In Fig. \ref{Fig1}c, we also compared the theoretical (black lines) and observed (colorful lines) equal-metallicity lines on the Petersen diagram. The lines for ${\rm [Fe/H]} = -1.5$ dex and $-1.8$ dex are in perfect agreement, but the theoretical line for ${\rm [Fe/H]} = -1.28$ dex corresponds to a smaller period ratio. We suspected that the possible reason for the discrepancy is the enhanced mass of the metal-rich RRd stars. Mass enhancement leads to an increase in the period ratio and a moderate decrease in the fundamental period\cite{Coppola15,Szabo2004}. Our observed period--period ratio--metallicity relation can help to optimize the theoretical models, mainly at the metal-rich and metal-poor ends, and the accuracy of the period ratios. Once the theoretical model can predict the period--period ratio--metallicity relation well, it can also predict the period--luminosity relation (PLR) of RRd stars. 

\begin{equation}\label{equ11}
\begin{split}
  {\rm [Fe/H]_{SDSS}} = (-10.33\pm0.47) \log P_{\rm F}+(-4.87 \pm 0.14),  \sigma=0.17 \, {\rm dex}, R^2=0.71, \\
  {\rm [Fe/H]_{SDSS}} = (-10.15\pm0.46) \log P_{\rm 1O}+(-6.11 \pm 0.19),  \sigma=0.17 \, {\rm dex}, R^2=0.71
\end{split}
\end{equation}

\begin{figure}
\begin{center}
\includegraphics[width=\hsize]{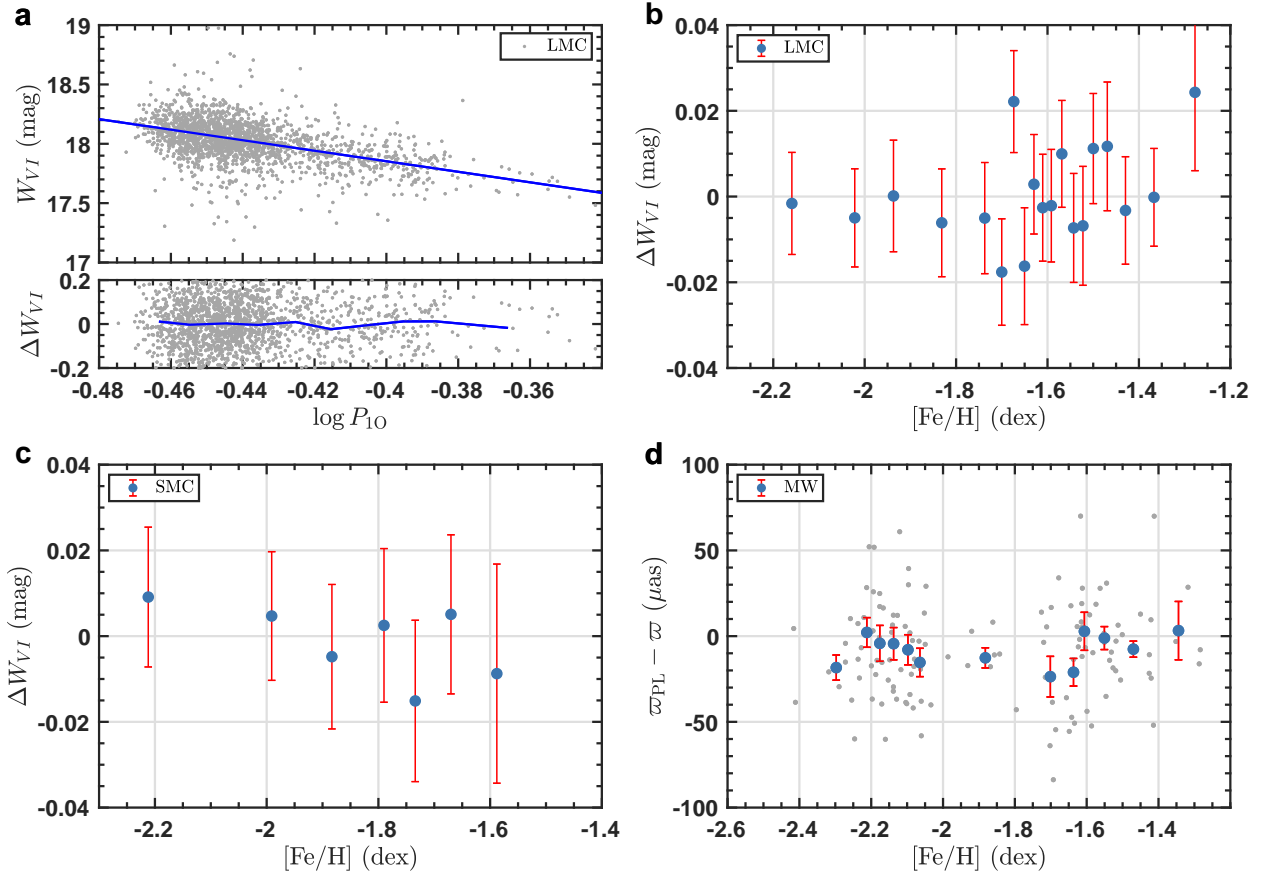}
\caption{\label{Fig2}{\bf Period--luminosity relation (PLR) and metallicity dependence examination for RRd stars.} {\bf a}, $W_{VI}-\log P_{\rm 1O}$ PLR based on LMC RRd stars. The light gray dots indicate the positions of LMC RRd stars, and the blue solid line is the best-fit PLR. The residual plot is shown at the bottom of this panel. {\bf b} and {\bf c}, The effect of metallicity on PLR residual $\Delta W_{VI}$ is examined by LMC and SMC RRd stars. The blue filled circles indicate the mean residual of each 100 RRd stars, binned in order of the metallicity. {\bf d}, The effect of metallicity on the zero point of $W_{G,BP,RP}-\log P_{\rm 1O}$ PLR is examined by the Milky Way's RRd stars (light gray dots) with good Gaia DR3 parallaxes. The PLR zero-point is converted into parallax $\varpi_{\rm PL}$ by assuming an LMC distance modulus of 18.48 mag. The blue filled circles denote the mean values of each 10 RRd stars, binned in order of the metallicity. The red error bars in panels {\bf b, c, d} denote the $1\sigma$ standard deviation in each bin. All of panels {\bf b, c, d} show that RRd star's PLRs are not depended on metallicity.}
\end{center}
\end{figure}

Fundamental-mode (RRab) and first overtone-mode (RRc) RR Lyr stars satisfy the period--metallicity--luminosity (PLZ) relations $M=a_0+a_1\log P+a_2{\rm [Fe/H]}$ in observations. In the optical band, these PLZ relations can be simplified to the metallicity--luminosity relations\cite{Catelan2015}. These relations can be predicted theoretically by combining the horizontal-branch evolutionary model and the pulsation model\cite{Marconi15}. The two periods of RRd stars also satisfy the PLZ relation of RRab stars and RRc stars, respectively. Combining the period--period ratio--metallicity relation, we obtained $M=b_0+b_1\log P_{\rm 1O}+b_2\log {P_{\rm 1O}/P_{\rm F}}$. Due to the approximate linear correlation between $\log P_{\rm 1O}$ and $\log {P_{\rm 1O}/P_{\rm F}}$, the relation can be simplified to $M=c_0+c_1\log P_{\rm 1O}$ or $M=d_0+d_1\log P_{\rm F}$. The reasons why we prefer to use the PLR rather than the period--period ratio--luminosity relation are discussed in the Methods section. We used the LMC RRd stars from the OGLE database\cite{Soszynski2019} to determine their PLRs. Since the distance of each RRd star with respect to the LMC mid-plane is non-negligible, using about 2000 RRd stars avoids the bias due to incompleteness. We adopted Wesenheit magnitude\cite{Madore1982} to reduce the effect of extinction, i.e. $W_{VI}=I-1.55(V-I)$. As for the Gaia passbands, the Wesenheit magnitudes is $W_{G, BP, RP}= G -1.90( BP - RP)$\cite{Ripepi2019}. The determined $M_W - \log P$ relations are shown in Equ. \ref{equ2} and Fig. \ref{Fig2}a. ${\rm DM_{LMC}}$ is the distance modulus of LMC. Since the primary period of an RRd star is the first-overtone period, it is preferable to use the first-overtone period to calculate the absolute magnitude, especially if the second period cannot be measured accurately. Nevertheless, the difference between the absolute magnitudes estimated by $P_{\rm F}$ and $P_{\rm 1O}$ is negligible for Gaia or OGLE RRd stars ($0.000\pm0.001$ mag).

\begin{equation}\label{equ2}
 \begin{aligned}
  &M_{W_{VI}} = (-4.523\pm0.156) \log P_{\rm F}+(16.620 \pm 0.048) - {\rm DM_{LMC}}, \sigma=0.132 \,{\rm mag}, \\
  &M_{W_{VI}} = (-4.434\pm0.153) \log P_{\rm 1O}+(16.079 \pm 0.067) - {\rm DM_{LMC}}, \sigma=0.132 \,{\rm mag}, \\
  &M_{W_{G, BP, RP}} = (-3.623\pm0.229) \log P_{\rm F}+(17.042 \pm 0.071) - {\rm DM_{LMC}}, \sigma=0.159 \,{\rm mag}, \\
  &M_{W_{G, BP, RP}} = (-3.557\pm0.225) \log P_{\rm 1O}+(16.606 \pm 0.098) - {\rm DM_{LMC}}, \sigma=0.159 \,{\rm mag}
 \end{aligned} 
\end{equation}

In addition to the theoretical derivation, there are direct observational evidences that can show that the PLR of RRd stars is independent of metallicity. The most direct evidence is that the PLR of RRd stars can be derived from the PLZ relation of RRab or RRc stars by simply using the period--metallicity relation of RRd stars to remove the metallicity dependence. 

LMC RR Lyr stars from the OGLE database were used here. We took RRc stars as an example and determined their PLR as $M_{W_{VI}} = (-3.14\pm0.03) \log P_{\rm 1O}+(16.66 \pm 0.01) - {\rm DM_{LMC}}, \sigma=0.136$ mag. The slope and intercept of RRc stars' PLR are very different from those of RRd stars (see Equ. \ref{equ2}). We then added a metallicity-dependent term $+0.13({\rm [Fe/H]}+1.64)$ to the PLR of RRc stars, where 0.13 comes from theoretical calculations\cite{Marconi15} and $-1.64\pm0.20$ dex is the mean metallicity of LMC RRd stars (calculated by Equ. \ref{equ1}). Here we assumed that RRc stars and RRd stars have the same mean metallicity. In the Milky Way, the mean metallicities of RRab, RRc, and RRd stars are $-1.50\pm0.37$ dex, $-1.62\pm0.42$ dex, and $-1.71\pm0.32$ dex, respectively, based on the Gaia DR3 RR Lyr sample and the SDSS metallicities. The mean metallicity differences between RRd stars and RRab, RRc stars are in the range of $0.1-0.2$ dex, and these differences have been considered since we used an error of 0.2 dex for the mean metallicity of RRd stars. This metallicity difference causes a $0.01-0.03$ mag change in the zero point of the PLR, but does not affect the slope. Combining the period--metallicity relation of RRd stars, we obtained the PLR as $M_{W_{VI}} = (-4.46\pm0.07) \log P_{\rm 1O}+(16.08 \pm 0.04) - {\rm DM_{LMC}}$, which is exactly consistent with the PLR of RRd stars. Similarly, we used RRab stars\cite{Marconi15,Neeley2019} to obtain the PLR of the fundamental mode as $M_{W_{VI}} = (-4.58\pm0.07) \log P_{\rm F}+(16.67 \pm 0.04) - {\rm DM_{LMC}}$. If we considered the coefficient difference in the logarithmic periods between the two PLRs as the possible remaining metallicity dependence of the PLRs. The results are $\Delta M_{W_{VI}}=(0.003\pm0.017){\rm [Fe/H]}$ and $\Delta M_{W_{VI}}=(0.006\pm0.017){\rm [Fe/H]}$ for PLRs based on the first-overtone and fundamental period, which are negligible. These consistencies indicate that calculating the luminosity using the PLR of RRd stars is equivalent to calculating the luminosity using the PLZ relation of RRab or RRc stars. In turn, the period--metallicity relation of RRd stars can be determined by combining the PLZ relations of RRab or RRc stars with the PLR of RRd stars, even without knowing the metallicity of any RRd stars (see Methods). This means that RRd stars are RRab or RRc stars that satisfy the period--metallicity relation.

We estimated the metallicity of RRd stars from the period--period ratio--metallicity relation (Equ. \ref{equ1}) and checked the correlation between the magnitude residual of PLR and the metallicity by LMC, SMC, and Galactic RRd stars separately. Note that this metallicity is not a completely independent measurement, but it can be used to check whether the period ratio affects the PLR through metallicity. For LMC and SMC, we binned RRd stars in order of metallicity, with a bin size of 100. This bin size allows us to detect deviations as low as 0.01 mag. We then fixed the slope of the PLR and calculated the mean magnitude residual $\Delta W_{VI} = W_{VI} - W_{VI,{\rm PL}}$ in each bin separately. In both LMC and SMC, we found that the correlation between the magnitude residual of the PLR and the metallicity is less than 0.03 mag/dex (see Fig. \ref{Fig2}b, c).

We estimated the PLR distances of 126 Galactic RRd stars with good Gaia DR3\cite{GaiaEDR3} parallaxes ($\varpi>0$, $\sigma_\varpi/\varpi<0.25$, $W_{G, BP, RP}<14$ mag and RUWE $< 1.4$, $\varpi$ denotes the Gaia DR3 parallax) by assuming an LMC distance modulus of 18.48 mag\cite{Pietrzynski2019}. The PLR distances were converted to parallaxes and we compared them with the Gaia DR3-corrected parallaxes $\varpi_{\rm corr}$\cite{Lindegren2021}. The mean error of Gaia parallax for this sample is 12\%. We found that the difference between the two parallaxes ${\rm zp}=\varpi-\varpi_{\rm PL}$ shows no correlation with metallicities (see Fig. \ref{Fig2}d). The mean parallax offset is ${\rm zp}=8.3\pm2.5\pm2.6$ $\mu {\rm as}$. The statistical error is the standard deviation divided by the root of the sample size, while the systematic error is propagated from the distance uncertainty (1.1\%) of LMC. This parallax offset agrees well with the result determined by classical Cepheids (${\rm zp}=14\pm6$ $\mu {\rm as}$\cite{Riess2021}), contact binaries (${\rm zp}=4.2\pm1.9$ $\mu {\rm as}$\cite{Ren2021}), and red giants (${\rm zp}=15\pm5$ $\mu {\rm as}$\cite{Zinn2019}). 

As a distance tracer, we used both the LMC distance and Gaia parallaxes to optimize the PLR of RRd stars. The PLR can then be used to determine the distances of distant galaxies or dwarf galaxies. The Gaia parallaxes can also provide an independent constraint for the PLR zero point of RRd stars. Based on the method of ref.\cite{Riess2021}, we used nonlinear least squares to fit Equ. \ref{equ3}. We fixed the slope of PLR $a_0=-3.557\pm0.225$ (see Equ. \ref{equ2}) and determined $a_1=M_{W_{G, BP, RP}}(P_{\rm 1O}=0.37 {\rm d})=-0.388\pm0.051$ mag and ${\rm zp}=13.4\pm5.8$ $\mu {\rm as}$. Based on the RRd stars with different parallaxes, the degeneracy of these two parameters is largely broken. Based on this PLR, the determined distance modulus of the LMC is $18.530\pm0.051$ mag.
By calculating the weighted average of the PLR zero points determined based on the Gaia parallax and the LMC distance ($M_{W_{G, BP, RP}}(P_{\rm 1O}=0.37 {\rm d})=-0.338\pm0.024$ mag), we obtained the final zero points of PLRs as $M_{W_{G, BP, RP}}(P_{\rm 1O}=0.37 {\rm d})=-0.348\pm0.022$ mag and $M_{W_{VI}}(P_{\rm 1O}=0.37 {\rm d})=-0.496\pm0.022$ mag (1.0\% distance uncertainty). Based on 617 SMC RRd stars, the determined distance modulus and average metallicity are ${\rm DM(SMC)}=18.913\pm0.007\pm0.022$ mag ($60.62\pm0.64$ kpc) and ${\rm [Fe/H]}=-1.87\pm0.19$ dex. Due to the existence of non-negligible irregular spatial structure of SMC\cite{Scowcroft2016,Ripepi2017}, the actual error of the average distance will be larger. The metallicity dispersion of SMC RRd stars is smaller than that of the LMC (0.21) and the Milky Way (0.24).

\begin{equation}\label{equ3}
\varpi_{\rm corr}=10^{-(0.2\times(W_{G, BP, RP}-a_0\times(\log P_{\rm 1O}-\log 0.37)-a_1)-2)}+{\rm zp}
\end{equation}

Compared to RRab stars, distance measurements based on RRd stars are no longer affected by metallicities. The difficult-to-measure metallicity is replaced by an easy-to-measure period. In addition, RRd stars can provide the metallicity distribution of the galaxy's old populations to help RRab stars' distance measurements. This is crucial because for most galaxies with RRab stars, we cannot directly measure the metallicities of RRab stars. RRab stars' metallicities are also difficult to infer from the average metallicity of the host galaxy. Although the metallicity of RRab star can be obtained indirectly from the light-curve shape through the parameter $\phi_{31}$\cite{Jurcsik1996}, this requires high-precision and high-sampling data, otherwise the prevalence of light-curve modulation in RRab will make $\phi_{31}$ measurements inaccurate. In the Gaia DR3 RR Lyr sample, the $\phi_{31}$-based metallicity error (only propagation error) is nearly 50 times larger than the $(P_{\rm 1O},P_{\rm 1O}/P_{\rm F})$-based metallicity error for the same number of epochs (see Extended Data Fig. 2). Besides, optimizing the $L({\rm RRab})=f(P, \phi_{31})$ relationship still requires more work because the value of $\phi_{31}$ is different in different bands. In contrast, RRd stars require only periods to estimate distances, which is not only convenient but also reduces systematic biases. The metallicity effect introduces a systematic error of about 1.8\% to the RRab stars-based distance measurements\cite{Tran2022}. The metallicity effect and the PLR zero point are the two most significant components of the systematic error in distance measurements of RR Lyr stars and Cepheids. We calculated the root sum square of these two errors and call it the base systematic error. The base systematic error is the lower limit error of the tracer in distance measurement, so it can be used as a criterion to evaluate the goodness of the distance tracer. When using RRab and RRd stars to measure the distances of galaxies or dwarf galaxies, the base systematic errors are $\sqrt{1.8\%^2+1.0\%^2}=2.1\%$ and $\sqrt{0.0\%^2+1.0\%^2}=1.0\%$. Compared to classical Cepheids, distance measurements based on RRd stars avoid the effects of metallicity and binarity. The fraction of RR Lyr star in a binary system is as low as 7\%\cite{Kervella2019}. Cepheid pulsations in both fundamental and first-overtone modes was also found to fulfill a period ratio--metallicity relation\cite{Kovtyukh2016}. However, the smaller number (95 in LMC) and the larger dispersion on the Petersen diagram limit the application of these Cepheids in distance measurements. The dispersion of the period ratios based on the quadratic curve fit is 0.0004 (RRd stars) and 0.0043 (Cepheids with F/1O modes).

\begin{figure}
\begin{center}
\includegraphics[width=\hsize]{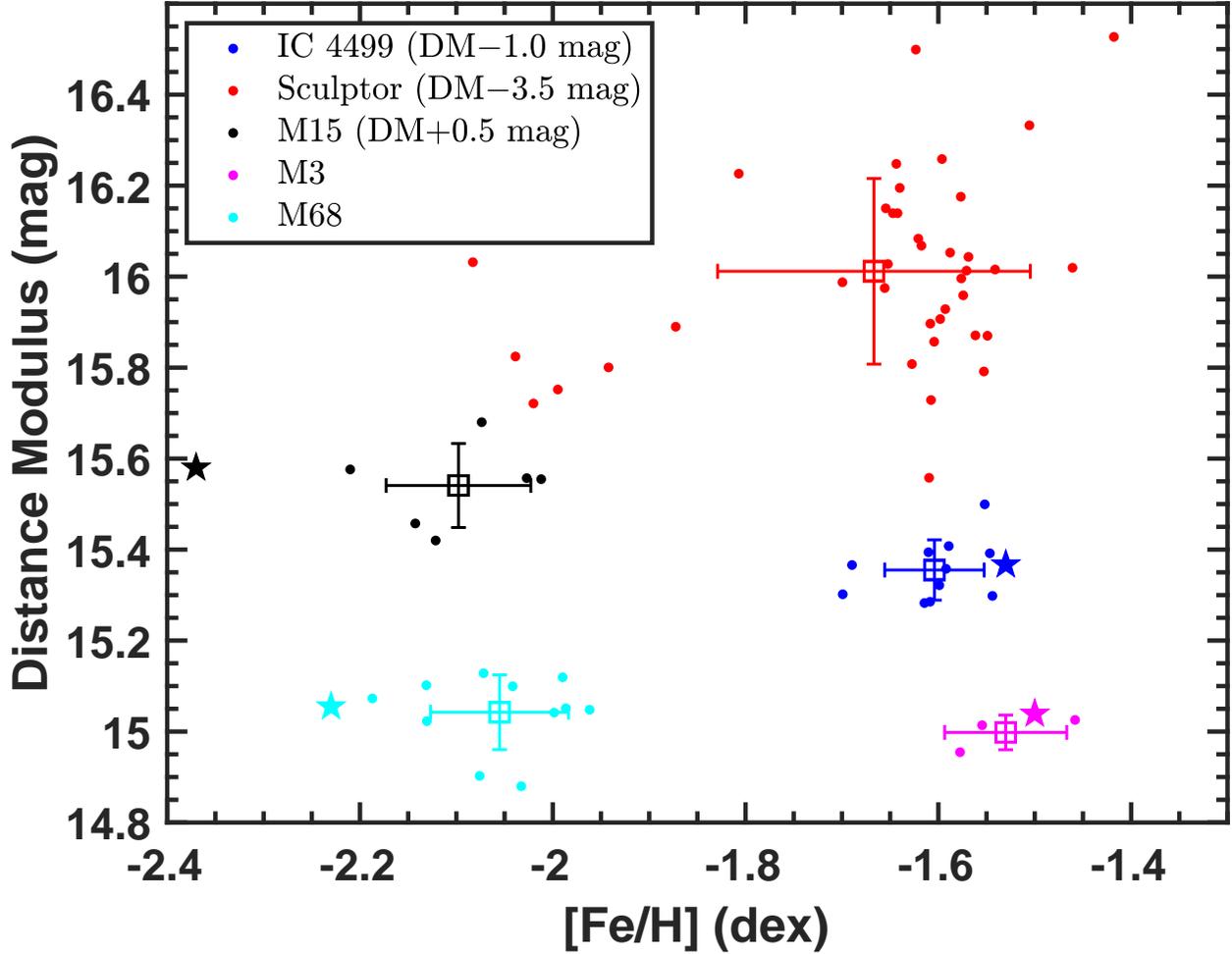}
\caption{\label{Fig3}{\bf Metallicity and distance modulus (DM) determination based on RRd stars.} The different colored dots indicate individual metallicity and distance modulus determinations based on RRd stars in IC 4499 (blue), Sculptor (red), M15 (black), M3 (magenta), and M68 (cyan). The mean values of RRd stars in five targets are shown as squares, while the $1\sigma$ internal errors are added. Solid pentagrams indicate the parameters of the four globular clusters, from the Harris's globular cluster catalog. Squares and pentagrams are colored the same as the dots. To make the figure clearer, the distance modulus of IC 4499 is subtracted by 1.0 mag, Sculptor is subtracted by 3.5 mag, and M15 is increased by 0.5 mag.}
\end{center}
\end{figure}

 We used RRd stars to study four globular clusters: IC 4499, M15, M3 and M68, a dwarf galaxy: Sculptor as examples. IC 4499, M15, and Sculptor are the three targets in the Gaia DR3 RR Lyr catalog with more than five RRd stars, with numbers of 11, 6, and 40, respectively. M3 and M68 contain 7 and 9 RRd stars with periods and $VI$ mean magnitude information from the literature\cite{Kains2015,Jurcsik2015}. The determined distances and metallicities based on RRd stars agree well with previous works\cite{Harris1996,Tran2022,Majewski1999} in $1\sigma$ and $2\sigma$ (see Fig. \ref{Fig3}), except for M15's metallicity ($2.9\sigma$) if we considered a metallicity uncertainty of 0.05 dex in literature. Our overall distance uncertainty is at the level of $2-3\%$ for globular clusters. The distance error of these four globular clusters determined by the different works is typically around 5\%, while the best distance error is around 2-3\%\cite{Baumgardt2021}. The independent distances provided by RRd stars can help to optimize the average distances obtained by combining different methods. The advantage is that it has a base systematic error of only 1\%. When combined with other distances (especially photometric distances), more components of the error can be eliminated. The metallicity scatters of RRd stars in globular clusters and the dwarf galaxy Sculptor are about 0.05-0.08 dex and 0.16 dex. Note that dispersion smaller than 0.16 dex may be unrealistic, and here it is certain that the dispersion of the period and the period ratio of RRd stars in globular clusters is much smaller than in dwarf galaxies.

For Sculptor, there are two groups of RRd stars with ${\rm [Fe/H]}=-1.59\pm0.06$ dex and ${\rm [Fe/H]}=-1.96\pm0.09$ dex, respectively. This is consistent with the bimodal distribution of metallicity inferred based on the horizontal branching discontinuity and the double red giant branch bumps\cite{Majewski1999}. The average metallicities of Sculptor based on 107 RR Lyr stars\cite{Clementini05} are $-1.83\pm0.26$ dex (on the scale of Ref.\cite{Zinn84}) and $-1.64\pm0.27$ dex (on the scale of Ref.\cite{Carretta97}), the latter being more consistent with our average metallicity $-1.67\pm0.16$ dex. There are 6 RRd stars among these RR Lyr stars. We found that our metallicity measurements are also consistent with the literature, with deviations of $-0.13 \pm0.25$ dex and $0.07 \pm0.26$ dex for two different scales.  The determined distance modulus of Sculptor is $19.51\pm0.03\pm0.02$ mag ($79.86\pm1.46$ kpc).
We find the mean metallicities of IC 4499 and M15 are ${\rm [Fe/H]}=-1.60\pm0.05$ dex and ${\rm [Fe/H]}=-2.10\pm0.08$ dex. Our M15 metallicity is 0.27 dex higher than the literature, but the deviation is still within $3\sigma$. The reason for the overestimation may be that the metallicity ${\rm [Fe/H]}=-2.37$ has reached the metal-poor end of our period--period ratio--metallicity relation. The determined distance moduli are ${\rm DM(IC 4499)}=16.36\pm0.05\pm0.03$ mag ($18.66\pm0.49$ kpc) and ${\rm DM(M15)}=15.04\pm0.06\pm0.02$ mag ($10.19\pm0.32$ kpc). 

The period ratios of RRd stars in M3 and M68 are not accurate enough to be used to determine metallicities. We adopted the first-overtone period and converted it to the period ratio based on an empirical quadratic function (Equ. \ref{equ4} in Methods). Metallicity is then determined by Equ. \ref{equ1}. Metallicities of M3 and M68 are  ${\rm [Fe/H]}=-1.53\pm0.06$ dex and ${\rm [Fe/H]}=-2.06\pm0.07$ dex. Note that four M3 RRd stars are not classical RRd stars and we excluded them in determining metallicity and distance (see Methods and discussion in Ref.\cite{Jurcsik2015}). The determined distance moduli are ${\rm DM(M3)}=15.00\pm0.08\pm0.02$ mag ($9.99\pm0.37$ kpc) and ${\rm DM(M68)}=15.04\pm0.04\pm0.02$ mag ($10.20\pm0.21$ kpc). 

The percentage of RRd in RR Lyr stars is 10\%, 5\% and $\sim3\%$ in SMC, LMC and the Milky Way halo, respectively. With the continued observations of the ZTF, OGLE and Gaia, the number of RRd stars in the Milky Way will increase by $\sim4000$ in the next few years. To date, there are at least 22 galaxies or dwarf galaxies with more than 100 RR Lyr stars\cite{Martinez-Vazquez2019}, all of which are suitable for distance (1-2\% accuracy) and metallicity measurements using RRd stars. With RRd star's PLRs determined for the Dark Energy Camera and the Hubble Space Telescope, RRd stars can trace distances and metallicities to 300 kpc and 1 Mpc. With future's China Space Station Telescope (CSST) and the Vera C. Rubin Observatory Legacy Survey of Space and Time (LSST), RRd stars will provide an independent distance ladder for the near-field universe to examine the distance ladder based on the classical Cepheid\cite{Riess2022} or the tip of the red giant branch\cite{Freedman2019}.

\begin{methods}

\subsection{RRd sample}
To obtain a larger sample of RRd stars with metallicities, we used the light curve of ZTF DR14 to help identify them. We targeted on RRab or RRc stars in Gaia DR3 with SDSS or LAMOST metallicity, which may not be classified as RRd stars due to the insufficient number of photometry. In analyzing the ZTF light curve, we used the Lomb-Scargle algorithm\cite{Lomb1976,Scargle1982} to obtain the primary period and then fitted the light curve using a sixth-order Fourier function\cite{Chen2020}. Then we performed the Lomb-Scargle algorithm on the residual light curve after pre-whitening with the primary period to obtain the second period. Only periods with low false-alarm probability ${\rm FAP} < 0.001$ were considered to be real periods.  We also excluded periods with light-variation amplitudes less than 0.03 mag to avoid false signals or non-classical RRd stars. Candidates with period ratios between 0.72 and 0.76 were finally selected as RRd stars, and we found that all of these RRd stars have period ratios between 0.741 and 0.748 (see Fig. \ref{Fig1}c). In this way, we found 186 RRd stars with metallicity. We also found 17 RRd stars that were not classified as RR Lyr by Gaia DR3, but were classified as RR Lyr by the ZTF periodic variable catalog\cite{Chen2020}. The final sample size of RRd stars we used to determine the period--period ratio--metallicity relation was 303. 55 RRd stars have multi-epoch spectra, and we estimated and adopted their average metallicities. By comparing the period ratios obtained from Gaia and ZTF photometry, we found that the period ratio error of RRd stars is $2.5\times10^{-5}$ in observation.

RR Lyr stars with metallicities are a very good sample to analyze the proportion of RRd stars. These RR Lyr stars are relatively bright, and their second period is easily detected by time-series photometry. ZTF photometry is very suitable for analyzing the proportion of RRd stars because it covers the sky field of SDSS and LAMOST and has an average of $\sim500$ photometry over a four-year span. Based on ZTF photometry, we found that for 707 RR Lyr stars with both LAMOST and SDSS parameters, the number of RRd stars is 29 ($4.1\%$). For 3076 and 4305 RR Lyr stars with only SDSS and LAMOST parameters, the numbers of RRd stars are 168 and 92, respectively. The average proportion of RRd stars is $3.6\%$. This is consistent with the proportion of RRd stars calculated based on the OGLE database\cite{Soszynski2019} in the anti-galactic center direction ($3\%$) and is much higher than the proportion of RRd stars in the bulge ($<1\%$).

For the sample of 126 RRd stars with good Gaia parallaxes, 68 of them are from the Gaia DR3 RRd star's catalog, and 47 are confirmed by ZTF photometry. We also added 11 nearby RRd stars\cite{Kovacs2021}.
\subsection{Anomalous/peculiar RRd stars}
Among the double-mode RR Lyr stars, there are a small number of anomalous or peculiar RRd stars in addition to the classical RRd stars. The main differences\cite{Nemec2021} between anomalous or peculiar RRd stars and classical RRd stars are: 1. Anomalous or peculiar RRd stars are located above or below the sequence of classical RRd stars on the Petersen diagram. 2. The dominant mode of anomalous or peculiar RRd stars is the fundamental mode while the dominant mode of classical RRd stars is the first-overtone mode. 3. Anomalous or peculiar RRd stars usually have long-term amplitude modulation. The main difference between peculiar RRd stars and anomalous RRd stars is that the amplitude ratio of peculiar RRd stars is usually less than 0.05. The use of classical RRd star relations to calculate the metallicity and luminosity of anomalous or peculiar RRd stars usually results in large deviations due to deviations from the sequence of classical RRd stars. We checked that the RRd samples used to obtain the period--period ratio--metallicity relation, PLRs, and the Gaia parallax offset are all classical RRd stars. There are four anomalous or peculiar RRd stars in the M3 globular cluster\cite{Jurcsik2015}, and we excluded them from the metallicity and distance analysis. It should be noted that there are dozens of short-period classical RRd stars ($P_{\rm 1O}/{P_{\rm F}}<0.74$), which are found in the Galactic bulge\cite{Soszynski2019} and are not in our sample. Whether these short-period classical RRd stars satisfy the relation of classical RRd stars requires future confirmation based on spectral parameters.

\subsection{Metallicity}
The metallicities we adopted are determined by SEGUE Stellar Parameter Pipeline on SDSS spectra and determined by LAMOST Stellar Parameter pipeline on LAMOST spectra. Three RRd stars with LAMOST ${\rm [Fe/H]} > 0.0$ dex are excluded because their spectra do not match correctly with the template ($T_{\rm eff}>10000$ K). The metallicities of 303 RRd stars are distributed between -2.6 dex and -0.8 dex, where the average internal error of the metallicity is 0.08 dex. To obtain its external error, we compared the spectral parameters of LAMOST and SDSS with the high-resolution spectroscopic parameters of the Apache Point Observatory Galactic Evolution Experiment (APOGEE)\cite{Majewski2017}. We chose stars with metallicities ${\rm [Fe/H]} < -1.0$ dex, which are more consistent with our RRd stars of interest. We found that the external error of metal-poor stars is 0.03 dex larger than that of solar metallicity stars. For comparison, we used one-time $3\sigma$ clipping to remove the outliers (a rate of 1.5\%) and then calculated the standard deviation. For 586 and 5840 metal-poor stars with SDSS and LAMOST metallicity, the standard deviations are 0.158 dex and 0.189 dex, respectively. If we select the sample by the internal error $\sigma_{\rm [Fe/H]} <0.04$ dex, then the standard deviation for the remaining 262 metal-poor stars with SDSS metallicity is 0.131 dex. For the remaining 3883 metal-poor stars with LAMOST metallicity, the standard deviation is 0.177 dex. The external error of the low-resolution spectroscopic metallicity is slightly smaller than these standard deviations, which also include the APOGEE metallicity error. We compared the external error with the dispersion of the period--period ratio--metallicity relation (Equ. \ref{equ1}) and found that they are comparable for the SDSS RRd stars. This suggests that the dispersion of the period--period ratio--metallicity relation arises from errors in the low-resolution spectroscopic metallicities.

\subsection{Mass}
Mass is a very poorly studied parameter of RR Lyr stars. Due to the low percentage of RR Lyr binaries and their existence only in wide binaries (with orbital periods longer than 1000 days), no dynamical masses of typical RR Lyr stars are currently known. The analysis of RR Lyr stars' masses can only be based on evolutionary masses. For the RRd star, Ref. \cite{Marconi15} provides the equation $\log M/M_\odot=-0.85\pm0.05 -(2.8\pm0.3)\log P_{\rm 1O}/{P_{\rm F}}-(0.097\pm0.003)\log Z$ to calculate its evolutionary mass. For 303 RRd stars with metallicities, the determined masses are between 0.58 and 0.85 $M_\odot$. The mass range also agreed with the prediction from the other theoretical model\cite{Szabo2004}. We calculated a linear correlation coefficient of $R^2=0.9984$ between metallicities and logarithmic masses, and such a high correlation is due to the existence of the period ratio--metallicity relation. This correlation suggests that the period--period ratio--metallicity relation for RRd stars does not require the introduction of mass as an independent variable. There is no metallicity dependence in the PLR of RRd stars, and likewise no mass dependence. By calculation, we obtained that the correlation between the PLR residuals and the mass or metallicity is consistent, both being low to negligible.

\subsection{Period--Period Ratio--Metallicity relation}
We performed a nonlinear check of the period--period ratio--metallicity relation and period--metallicity relation. Based on the current 303 RRd stars, we found that the use of second- to fourth-order polynomials did not lead to a reduction in the root-mean-square error (RMSE) of the fit. We also tried more than 30 machine learning regression methods (e.g., support vector machines, Gaussian process regression, neural networks, etc.) to explore the complex relation between metallicity and period, period ratio. The results show that the linear regression learner has the smallest RMSE. We performed Markov-chain Monte Carlo (MCMC) simulations to estimate the coefficient error of ${\rm [Fe/H]}=e_0\log P_{\rm 1O}/{P_{\rm F}}+e_1\log P_{\rm 1O}+e_2$ and confirmed that the errors in Equ. \ref{equ1} were not underestimated. The period--period ratio--metallicity relation is applicable for RRd stars with $0.741\le P_{\rm 1O}/P_{\rm F} \le0.748$, which includes the majority of classical RRd stars. 

\subsection{Period--Metallicity relation}
The period--metallicity relation of RRd stars can be derived by combining the RR Lyr's PLZ relation and RRd star's PLR. We used RRab stars (from the OGLE LMC sample) as an example
and determined their PLR as $M_{W_{VI}} = (-3.03\pm0.02) \log P_{\rm F}+(17.15 \pm 0.01) - {\rm DM_{LMC}}, \sigma=0.133$ mag. We then added a metallicity-dependent term $+0.15({\rm [Fe/H]}+1.64)$ to the PLR of RRab stars, where 0.15 comes from the theoretical calculations\cite{Marconi15} and observations\cite{Neeley2019}. When RRab stars become RRd stars, they also follow the PLR of RRd stars (Equ. \ref{equ2}). Based on these two relations, we eliminated $M_{W_{VI}}$ and ${\rm DM_{LMC}}$ and derived the period--metallicity relation ${\rm [Fe/H]_{SDSS}} = (-9.96\pm1.04) \log P_{\rm F}+(-5.17 \pm 0.32)$ dex. Similarly, we used RRc stars to derive the period--metallicity relation for first-overtone mode as ${\rm [Fe/H]_{SDSS}} = (-9.92\pm1.20) \log P_{\rm 1O}+(-6.11 \pm 0.52)$ dex. These relations agree well with RRd stars' period--metallicity relations (Equ. \ref{equ11}). Therefore, RRd stars are RRab or RRc stars that satisfy the period--metallicity relation.

\subsection{Period--Period Ratio Relations}
The primary period determined from short time-span observation is more accurate than the second period. In this case, the period ratio of RRd stars can be better derived from the Petersen diagram. The primary period of the classical RRd star is the first-overtone period, and the conversion equation is given in Equ. \ref{equ4}. This relation was established by a quadratic curve fit to the OGLE LMC RRd stars. The period ratio error of 0.0004 is only one order of magnitude higher than the period ratio error of the ZTF or Gaia observations. The metallicity calculated with Equ. \ref{equ1} based on this approximate period ratio has a dispersion of 0.039 dex compared to the metallicity calculated with the observational period ratio.

\begin{equation}\label{equ4}
  P_{\rm 1O}/P_{\rm F} = (-0.6567\pm0.0200)\times P_{\rm 1O}^2+(0.5381\pm0.0151)\times P_{\rm 1O}+(0.6357\pm0.0029), \sigma=0.0004\\
\end{equation}

\subsection{Period--Luminosity Relation}
For the OGLE system, we excluded a few foreground RRd stars by $I>18.05$ mag (LMC) and $I>18.35$ mag (SMC), respectively, when obtaining the PLR. For the Gaia system, only $\sim 300$ RRd stars have a mean magnitude based on light-curve analysis. To avoid the PLR being biased by incompleteness, we complemented $\sim 800$ RRd stars ($\sigma_{\rm BP}<0.05$ mag and $\sigma_{\rm RP}<0.05$ mag) with Gaia DR3 statistical mean magnitude. However, considering that Gaia's statistical mean magnitude was based on more than 40 measurements, using this value would only increase the dispersion of the PLR (from 0.13 mag to 0.16 mag) without affecting the zero point. Based on future data released by Gaia, the dispersion of Gaia-based PLR will decrease.

\subsection{Zero point of Period--Luminosity Relation}
In the main text, we determined the zero point of RRd stars' PLR assuming a fixed slope from the LMC PLR. Here we also show the results using an unfixed slope, i.e. $M_{W_{G, BP, RP}}(P_{\rm 1O}=0.37 {\rm d})=-0.391\pm0.052$ mag, ${\rm zp}=13.5\pm5.9$, $a_0=-3.393\pm0.671$. We can see that the RRd star's PLRs based on the LMC distance and Gaia parallax agree well with each other. We finally use the PLR slope based on LMC RRd stars to establish the best PLR.

In the main text, we used a conservative PLR zero point of RRd stars that can be optimized in future work. If we assumed that there is no offset in the Gaia parallax, the PLR zero point obtained is $M_{W_{G, BP, RP}}(P_{\rm 1O}=0.37 {\rm d})=-0.286\pm0.022$ mag. Besides, based on the parallax offset of contact binaries, we determined a zero point of $M_{W_{G, BP, RP}}(P_{\rm 1O}=0.37 {\rm d})=-0.317\pm0.022$ mag. An assumption is introduced here that the sample of RRd stars and the sample of contact binaries have the same mean Gaia parallax offset. Since the Gaia parallax offset is related to the spatial position, color and $G$-band apparent magnitude\cite{Lindegren2021}, the assumption holds only if these two samples have the same mean value in these parameters. Based on the parallax offsets calculated from a fine selection of other tracers, or the future Gaia DR4 parallax, the zero-point uncertainty of the RRd stars' PLR can be improved to 0.022 mag. When combined with the LMC-based zero-point $M_{W_{G, BP, RP}}(P_{\rm 1O}=0.37 {\rm d})=-0.338\pm0.024$ mag, a zero-point uncertainty of 0.016 mag can be obtained (0.7\% distance uncertainty).

\subsection{Period--Period ratio--Luminosity Relation}
We obtained the period--period ratio--luminosity relations for RRd stars using the same samples as for PLRs (Equ. \ref{equ5}). We found that their RMSEs are not optimized compared to PLRs. By analyzing the PLR residuals $\Delta M_W$, we found that the period--period ratio--luminosity relation may be slightly overfitted at the long-period end and the small period ratio end. For the LMC RRd stars, the absolute magnitude deviations calculated with these two relations are $\Delta M_{W_{VI}} = 0.000\pm0.008$ mag and $\Delta M_{W_{G, BP, RP}} = 0.000\pm0.010$ mag, and the deviations are not significant compared to the RMSEs. This suggests that these two relations are consistent when using larger numbers of RRd stars to measure the distances of galaxies. As with metallicity, a linear regression relation between absolute magnitude and period or period ratio is the most appropriate, out of 30 machine learning regression methods.

\begin{equation}\label{equ5}
 \begin{aligned}
  M_{W_{VI}} =& (-3.800\pm0.283) (\log P_{\rm 1O}-\log 0.37) + (-30.06\pm8.58)(\log \frac{P_{\rm 1O}}{P_{\rm F}}-\log0.745)\\
  &+(17.987 \pm 0.004) - {\rm DM_{LMC}}, \sigma=0.132 \,{\rm mag}, \\
  M_{W_{G, BP, RP}} = & (-2.855\pm0.399) (\log P_{\rm 1O}-\log 0.37) + (-34.83\pm16.45)(\log \frac{P_{\rm 1O}}{P_{\rm F}}-\log0.745)\\
  &+(18.135 \pm 0.006) - {\rm DM_{LMC}}, \sigma=0.159 \,{\rm mag}, \\
 \end{aligned} 
\end{equation}
\subsection{Distance Uncertainties}
The statistical uncertainty was estimated by $\sigma_{\rm stat}=\max(\sigma,\sigma_{\rm PL})/\sqrt{n}$. $\sigma$ is the standard deviation of RRd stars' distance moduli in each globular cluster or dwarf galaxy, while n denotes the number of used RRd stars. $\sigma_{\rm PL}$ is the RMSE of PLR. Normally, $\sigma$ will be greater than or equal to $\sigma_{\rm PL}$. However, when the sample size is too small, $\sigma$ can be smaller than $\sigma_{\rm PL}$. We conservatively took their maximum values.  The systematic uncertainties include those caused by the PLR zero point, extinction, and metallicity. The PLR zero point was based on LMC distance and the Gaia DR3 parallaxes with an uncertainty of $\sigma_{\rm ZP}=1.0\%$. The extinction uncertainty was estimated by assuming an $\sigma_{\rm ext, coef}=5\%$ uncertainty in the coefficient of Wesenheit magnitude. The PLR and distance of RRd stars were re-determined based on the new coefficient and the distance modulus difference was assumed as the extinction uncertainty $\sigma_{\rm ext}$. IC4499 has an extinction uncertainty of $\sigma_{\rm ext}=0.02$ mag, while the other five targets (including SMC) have $\sigma_{\rm ext}<0.01$ mag.  The effect of the metallicity on the PLR is negligible. The final systematic uncertainty was estimated by $\sigma_{\rm syst}=\sqrt{\sigma_{\rm ZP}^2+\sigma_{\rm ext}^2}$.

\end{methods}

\subsection{Data Availability}
The full data set of the RRd sample used to determine the period-period ratio-metallicity relation is available in Supplementary Data 1. The full data set of the RRd sample used to determine the offset of Gaia parallax is available in Supplementary Data 2.
The data supporting the plots in this paper and other results from
this study are available from the corresponding author upon reasonable
request.

\subsection{Code Availability}
The MATLAB codes used in this study are available from the corresponding author upon reasonable request.

\begin{addendum}
 \item [Acknowledgements] 
   X.C. acknowledge support from the National Key Research and development Program of China (2022YFF0503404). X.C.,  S.W. and L.D. thank support from the National Natural Science Foundation of China through grants 12173047,  12003046, 12233007, 12133002, 11903045 and 11973001. X.C. and S.W. acknowledge support from the Youth Innovation Promotion Association of the Chinese Academy of Sciences (2022055 and 2023065). L.D. thank support from Major Science and Technology Project of Qinghai Province (2019-ZJ-A10). S.W. thank the support from the National Key Research and development Program of China (2019YFA0405504). We used of data from the European Space Agency mission Gaia (https://www.cosmos.esa.int/gaia), processed by the Gaia Data Processing and Analysis Consortium (DPAC, https://www.cosmos.esa.int/web/gaia/ dpac/consortium). Funding for the DPAC has been provided by national institutions, in particular the institutions participating in the Gaia Multilateral Agreement. This work has made use of LAMOST and SDSS data.  Guoshoujing Telescope (the Large Sky Area Multi-Object Fiber Spectroscopic Telescope, LAMOST) is a National Major Scientific Project built by the Chinese Academy of Sciences. Funding for the Sloan Digital Sky Survey IV has been provided by the Alfred P. Sloan Foundation, the U.S. Department of Energy Office of Science, and the Participating Institutions. SDSS-IV acknowledges support and resources from the Center for High-Performance Computing at the University of Utah. We used observations obtained with the Samuel Oschin 48-inch Telescope at the Palomar Observatory as part of the Zwicky Transient Facility project. ZTF is supported by the National Science Foundation under grant AST-1440341.

 \item [Author contributions] 
   X.C. contributed to project designing, data preparation, analysis and manuscript writing. 
   J.Z. contributed to data analysis and manuscript writing. 
   S.W. engaged in scientific discussions and contributed to manuscript writing. 
   L.D. contributed research support. 
   All authors reviewed and commented on the final manuscript.

 \item[Competing Interests] The authors declare no competing interests.
 \item[Correspondence] Correspondence and requests for materials
   should be addressed to Xiaodian Chen (chenxiaodian@nao.cas.cn).
\end{addendum}

\clearpage

\end{document}